\documentclass[12pt]{article}
\usepackage{amssymb,amsmath,epsfig}

\begin{document}

\title{\bf Cosmological Analysis of Pilgrim Dark Energy in Loop Quantum Cosmology}
\author{Abdul Jawad \thanks{jawadab181@yahoo.com;~~abduljawad@ciitlahore.edu.pk}\\
Department of Mathematics, COMSATS Institute of\\ Information
Technology, Lahore-54000, Pakistan.}

\date{}

\maketitle
\begin{abstract}
The proposal of pilgrim dark energy is based on speculation that
phantom-like dark energy (with strong enough resistive force) can
prevent black hole formation in the universe. We explore this
phenomenon in loop quantum cosmology framework by taking Hubble
horizon as an infra-red cutoff in pilgrim dark energy. We evaluate
the cosmological parameters such as Hubble, equation of state
parameter, squared speed of sound and also cosmological planes like
$\omega_{\vartheta}-\omega'_{\vartheta}$ and $r-s$ on the basis of
pilgrim dark energy parameter ($u$) and interacting parameter
($d^2$). It is found that values of Hubble parameter lies in the
range $74^{+0.005}_{-0.005}$. It is mentioned here that equation
state parameter lies within the ranges $-1\mp0.00005$ for $u=2,~1$
and $(-1.12,-1),~(-5,-1)$ for $u=-1,-2$, respectively. Also,
$\omega_{\vartheta}-\omega'_{\vartheta}$ planes provide $\Lambda$CDM
limit, freezing and thawing regions for all cases of $u$. It is also
interesting to mention here that
$\omega_{\vartheta}-\omega'_{\vartheta}$ planes lie in the range
($\omega_{\vartheta}=-1.13^{+0.24}_{-0.25},~\omega'_{\vartheta}<1.32$).
In addition, $r-s$ planes also corresponds to $\Lambda$CDM for all
cases of $u$. Finally, it is remarked that all the above constraints
of cosmological parameters shows consistency with different
observational data like Planck, WP, BAO, $H_0$ and SNLS.
\end{abstract}

\section{Introduction}

The accelerated expansion of the universe is one of the biggest
achievements in the subject of cosmology \cite{1aa}. This expansion
phenomenon follows through mysterious form of force called dark
energy (DE). However, the nature of DE is still unknown. Different
researchers have tried to explore the nature of DE through various
aspects via theoretical and observational ways. As a result, they
proposed different dynamical DE models as well as modified theories
of gravity. The dynamical DE models have been developed in the
scenarios of general relativity and quantum gravity. The pioneer
candidate of DE is cosmological constant but it has two severe
problems \cite{1a}. As an alternative to this candidate, the
proposals of family of chaplygin gas \cite{2}, holographic
\cite{3,4}, new agegraphic \cite{5}, polytropic gas \cite{6},
pilgrim \cite{N5}-\cite{N7} DE models have been come forward.

The holographic DE (HDE) has become an attractive DE model nowadays,
which is developed in the context of quantum gravity and widely used
in solving the cosmological problems. The main idea of this model
has come from holographic principle which is stated as \textit{the
number of degrees of freedom of a physical system should scale with
its bounding area rather than its volume} \cite{suss}. With the help
of this principle, a relationship between ultraviolet and infrared
(IR) cutoffs has been proposed by suggesting that the size of a
system should not exceed the mass of black hole (BH) of the same
size \cite{cohen}. By using this relationship, Li \cite{4} developed
HDE density as follows
\begin{equation*}
\rho_{\Lambda}=3n^{2}M^2_{p}L^{-2},
\end{equation*}
here, $n,~M_{p},~L$ indicate HDE constant, the reduced Planck
constant, IR cutoff, respectively. On the basis of compatibility of
HDE with the present day observations, different IR cutoffs have
been proposed which includes Hubble, particle, event horizons,
conformal age of the universe, Ricci scalar, Granda-Oliveros and
higher derivative of Hubble parameter \cite{5a}-\cite{5c} etc.

According to Cohen et al. \cite{cohen}, the bound of energy density
from the idea of formation of BH in quantum gravity. However, it is
suggested formation of BH can be avoided through appropriate
repulsive force which resists the matter collapse phenomenon. This
force can only provide phantom DE in spite of other phases of DE
like vacuum and quintessence DE. By keeping in mind this phenomenon,
Wei \cite{N5} has suggested the DE model called pilgrim DE (PDE) on
the speculation that phantom DE possesses the large negative
pressure as compared to the quintessence DE which helps in violating
the null energy condition and possibly prevent the formation of BH.
In the past, many applications of phantom DE exist in the
literature. For instance, phantom DE is also play an important role
in the wormhole physics where the event horizon can be avoided due
to its presence \cite{shar}.

Also, it plays role in the reduction of mass due to its accretion
process onto BH. Many works have been done in this support through a
family of chaplygin gas \cite{30}. It was also argued in the context
of scalar field that BH area reduces up to 50 percent through
phantom scalar field accretion onto it \cite{9}. According to Sun
\cite{10}, mass of BH tends to zero when the universe approaches to
big rip singularity. It was also suggested that BHs might not be
exist in the universe in the presence of quintessence-like DE which
violates only strong energy condition \cite{11}. However, these
works do not correspond to reality because quintessence DE does not
contain enough resistive force to in order to avoid the formation of
BH.

The above discussion is motivated to Wei \cite{N5} in developing the
PDE model. He analyzed this model with Hubble horizon through
different theoretical as well as observational aspects. Also,
Saridakis et al. \cite{S1}-\cite{S9} have discussed the widely the
crossing of phantom divide line, quintom as well as phantom-like
nature of the universe in different frameworks and found interesting
results in this respect. Recently, we have investigated this model
by taking different IR cutoffs in flat as well as non-flat FRW
universe with different cosmological parameters as well as
cosmological planes \cite{N6,N7}. This model has also been
investigated in different modified gravities \cite{N47}-\cite{N49}.
In the present paper, we check the role of PDE in loop quantum
cosmology (LQC). We develop different cosmological parameters and
planes. The format of the paper is as follows. In the next section,
we provide the basic equations corresponding to PDE models. Also, we
discuss the Hubble parameter, EoS parameter and squared speed of
sound in section \textbf{3}. Section \textbf{4} explores
$\omega_{\vartheta}-\omega_{\vartheta}'$ as well as statefinders
planes. In the last section, we summarize our results.

\section{Loop Quantum Cosmology and Pilgrim Dark Energy}

Nowadays, the discussion of DE phenomenon has also been done widely
in the context of LQC to describe the quantum effects on our
universe. The LQC is an interesting and attractive application of
the Loop Quantum Gravity in the cosmological framework and it
possesses the properties of non-perturbative and background
independent quantization of gravity \cite{L22}-\cite{L27}. In recent
years, many DE models have been studied in the scenario of LQC
\cite{L27a,L27b}. Jamil et al. \cite{L29} have explored cosmic
coincidence problem phenomenon of modern cosmology by taking
modified chaplygin gas coupled to dark matter. Also, some authors
found that the future singularity appearing in the standard FRW
cosmology can be avoided by loop quantum effects \cite{L30}.
Chakraborty et al. \cite{L31} have made observational study of
modified chaplygin gas in LQC. Here, we develop basic scenario of
interacting PDE (with Hubble horzion) with cold dark matter (CDM) in
LQC. The equation of motion in LQC has the form
\begin{equation}\label{1}
H^2=\frac{1}{3m^2_{pl}}\rho\left(1-\frac{\rho}{\rho_c}\right),
\end{equation}
where $\rho$ indicates the sum of CDM and PDE densities. Also,
$\rho_{c}=\frac{\sqrt{3}}{16\pi^2\gamma^3G^2\hbar}$ represents the
critical loop quantum density, $\gamma$ appears as dimensionless
Barbero-Immirzi parameter. It is predicted that the big bang, big
rip and other future singularities at semi classical regime can be
avoided in LQC. Moreover, the modification in standard FRW cosmology
due to LQC becomes more dominant and the universe begins to bounce
and then oscillate forever.

It is argued that phantom DE with strong negative pressure can push
the universe towards the big rip singularity where all the physical
objects lose the gravitational bounds and finally dispersed. The PDE
model is also developed in the favor of this scenario which is
defined as
\begin{eqnarray}\label{pde}
\rho_{\vartheta}=3n^2 m_p^{4-u}L^{-u},
\end{eqnarray}
here $u$ represents the PDE parameter. Wei explored the PDE model
with different possible theoretical and observational ways to make
the BH free phantom universe with Hubble horizon ($L=H^{-1}$)
through PDE parameter.

In this work, we also choose PDE with the Hubble horizon which is
the pioneer IR cutoff. Initially, it is plagued with a problem that
its EoS parameter provides inconsistent behavior with present status
of the universe \cite{5a}. This deficiency has been settled down
with the passage of time by pointing out that HDE with this IR
cutoff can explain the present scenario of the universe in the
presence of interaction with DM \cite{20a}. Also, the results of
different cosmological parameters have been established through
different observational schemes by choosing HDE model with Hubble
scale \cite{20b,20c}. Sheykhi \cite{12a} has discussed this model by
taking interaction with CDM and pointed out that such model
possesses the ability to explain the present scenario of the
universe.

In this work, we take interaction between PDE with CDM which takes
the following form
\begin{eqnarray}\label{4}
\dot{\rho}_{m}+3H\rho_{m}=\Theta,\quad
\dot{\rho}_{\vartheta}+3H(\rho_{\vartheta}+p_{\vartheta})=-\Theta,
\end{eqnarray}
where $\Theta$ possesses dynamical nature and appears as interaction
term between CDM and PDE. Different forms of this interaction term
has been proposed out of which we use the following form
\begin{equation}\label{5}
\Theta=3d^2H\rho_{m},
\end{equation}
where $d^2$ is an interacting constant which appears as interaction
parameter and exchanges the energy between CDM and DE components.
This form of interaction term has been explored for energy transfer
through different cosmological constraints. The sign of coupling
constant decides the decay of energies either DE decays into CDM
(when the interacting parameter is positive) or CDM decays into DE
(when the interacting parameter is negative). The present analysis
from different aspects imply that the phenomenon of DE decays into
CDM which is more acceptable and favors the observational data.
Hence, the Eqs. (\ref{4}) and (\ref{5}) give
\begin{equation}\label{10}
\rho_{m}=\rho_{m0}a^{3(d^2-1)}.
\end{equation}
Also, by taking the differentiation of $\rho_{\vartheta}$ (with
Hubble horizon) with respect to $x=\ln a$, we get
\begin{equation}\label{10a}
\rho'_{\vartheta}=u\rho_{\vartheta}\frac{\dot{H}}{H^2}.
\end{equation}

\section{Cosmological Parameters in LQC}

In this section, we will discuss the physical significance of
cosmological parameters corresponding to PDE with Hubble horizon in
LQC scenario.

\subsection{Hubble Parameter}

In order to check the behavior of Hubble parameter in this
framework, we can find the following expression after some
calculations
\begin{equation}\label{7a}
\frac{\dot{H}}{H^2}=-\frac{\rho_m}{M^2_pH^2(a)}
\left[\frac{2\rho_c}{\rho_c-2(\rho_\vartheta+\rho_{m})}-\frac{u\rho_{\vartheta}}{3M^2_pH^2(a)}\right].
\end{equation}
\begin{figure} \centering
\epsfig{file=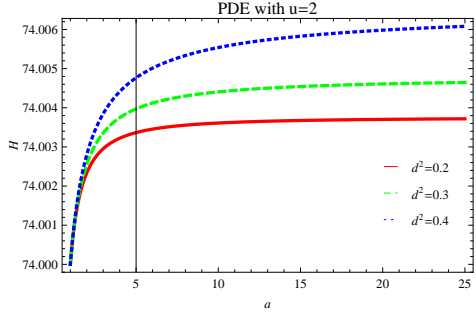,width=.50\linewidth}\caption{Plot of $H$ versus
$a$ for PDE in LQC with $u=2$.}
\end{figure}
\begin{figure} \centering
\epsfig{file=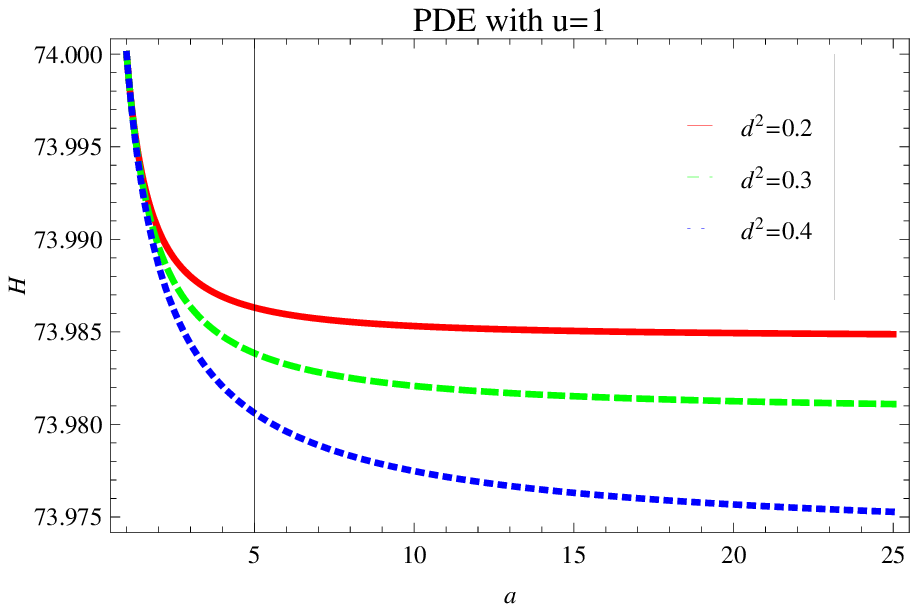,width=.50\linewidth}\caption{Plot of $H$ versus
$a$ for PDE in LQC with $u=1$.}
\end{figure}
\begin{figure} \centering
\epsfig{file=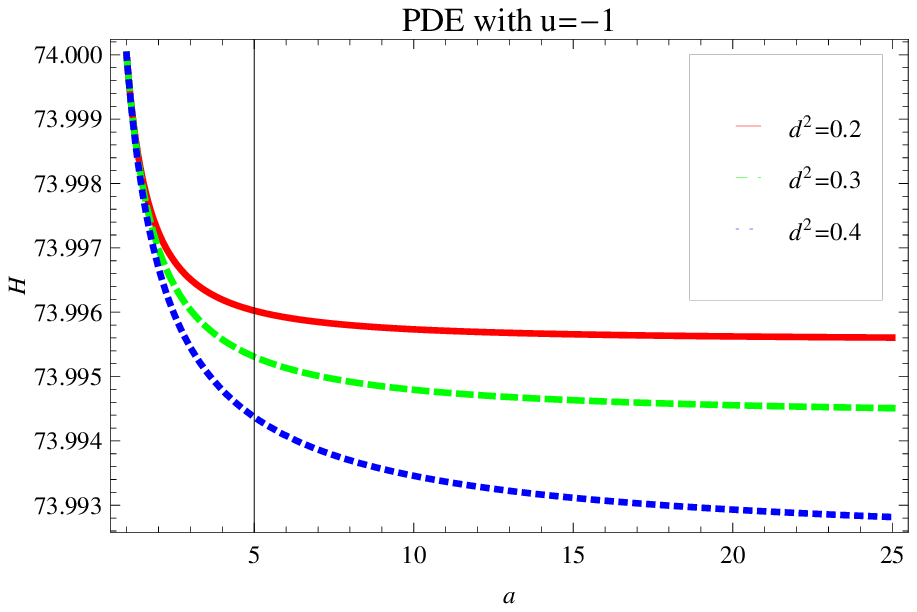,width=.50\linewidth}\caption{Plot of $H$ versus
$a$ for PDE in LQC with $u=-1$.}
\end{figure}
\begin{figure} \centering
\epsfig{file=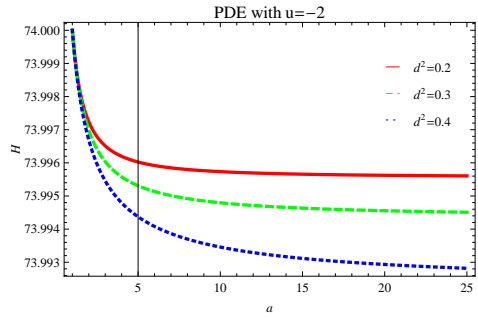,width=.50\linewidth}\caption{Plot of $H$ versus
$a$ in LQC with $u=-2$.}
\end{figure}

We solve the above differential equation (\ref{7a}) numerically by
using Eqs.(\ref{1})-(\ref{7a}) in terms of $H$ and plot it against
scale factor $a$ for four different values of $u=2,~1,-1,-2$ as
shown in Figures \textbf{1-4}. The initial condition of $H$ is taken
as $H[a_0]\simeq74$ as mentioned in Planck observations
\cite{Planck}. It has been greatly improved the precision of the
cosmic distance scale through two recent analysis. Riess et al.
\cite{Riess} use HST observations of Cepheid variables in the host
galaxies of eight SNe Ia to calibrate the supernova
magnitude-redshift relation. Their "best estimate" of the Hubble
constant, from fitting the calibrated SNe magnitude-redshift
relation, is
\begin{eqnarray*}
H_0&=&73.8\pm2.4km s^{-1}Mpc^{-1}~~~~~~~\text{(Cepheids+SNe Ia)},\\
\end{eqnarray*}
where the error is $1\sigma$ level and includes known sources of
systematic errors. Freedman et al. \cite{Freedman}, as part of the
Carnegie Hubble Program, use Spitzer Space Telescope mid-infrared
observations to recalibrate secondary distance methods used in the
HST Key Project. These authors find
\begin{eqnarray*}
H_0=[74.3 \pm 1.5\text{(statistical)} \pm 2.1 \text{(systematic)}]
km s^{-1} Mpc^{-1}\\\nonumber\text{(Carnegie HP)}.
\end{eqnarray*}

In the present paper, it can be observed that the trajectories of
Hubble parameter $H(a)$ attains the values approximately to
$74^{+0.005}_{-0.005}$. Hence, the present results of Hubble
parameter shows consistency with the above results found through
observations.

\subsection{The Equation of State Parameter}

By using all above equations, we can obtain the EoS parameter as
\begin{equation}\label{7b}
\omega_{\vartheta}=-1-d^2\frac{\rho_m}{\rho_\vartheta}+\frac{u\rho_m}{3H^2(a)}
\left[\frac{2\rho_c}{\rho_c-2(\rho_\vartheta+\rho_{m})}-\frac{u\rho_{\vartheta}}{3M^2_pH^2(a)}\right]
\end{equation}
The plots of EoS parameter versus $a$ are shown in Figures
\textbf{5-8} for four different values of $u$. In Figure \textbf{5}
($u=2$), the trajectory of $\omega_{\vartheta}$ starts from phantom
region and with the passage of time, it approaches to $\Lambda$CDM
limit for the interacting cases $d^2=0.2,~0.3$. However, it remains
in the phantom region for the interacting case $d^2=0.4$. In case of
$u=1$ (Figure \textbf{6}), the trajectories of EoS parameter remains
in quintessence region for $d^2=0.4$ while it approaches (from
quintessence region) to $\Lambda$CDM limit for the other two cases
of $d^2$. For $u=-1,-2$ (Figures \textbf{7-8}), the EoS starts from
phantom with comparatively high value and goes towards $\Lambda$CDM
limit for $d^2=0.2,~0.3$ and always remains in phantom for other
case of $d^2$. Moreover, the constraints on EoS parameter has been
put forward by Ade et al. \cite{Planck} (Planck data)
\begin{figure} \centering
\epsfig{file=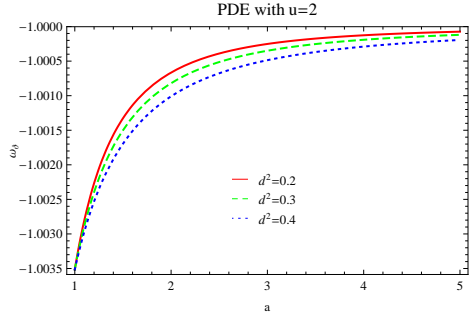,width=.50\linewidth}\caption{Plot of
$\omega_{\vartheta}$ versus $a$ for PDE in LQC with $u=2$.}
\end{figure}
\begin{figure} \centering
\epsfig{file=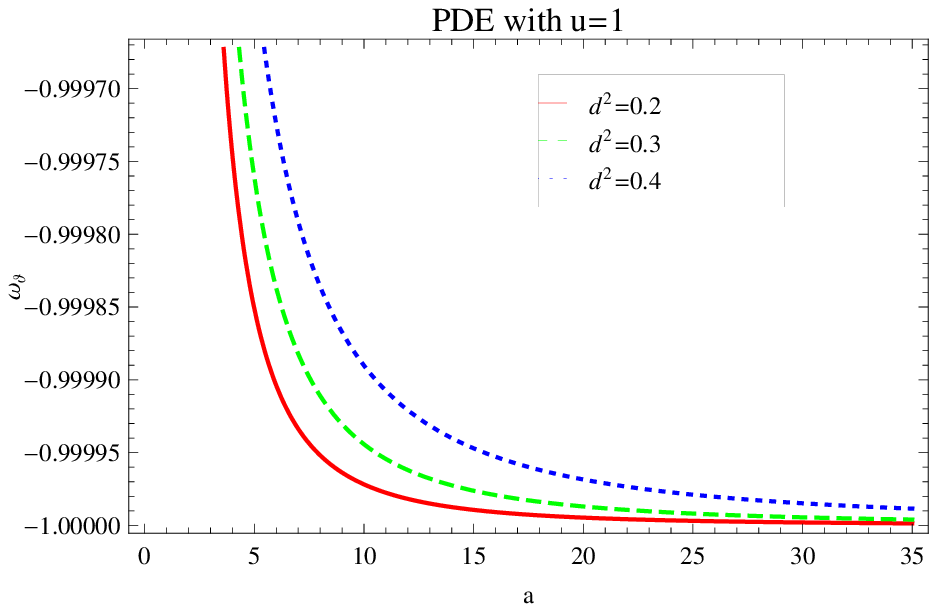,width=.50\linewidth}\caption{Plot of
$\omega_{\vartheta}$ versus $a$ for PDE in LQC with $u=1$.}
\end{figure}
\begin{figure} \centering
\epsfig{file=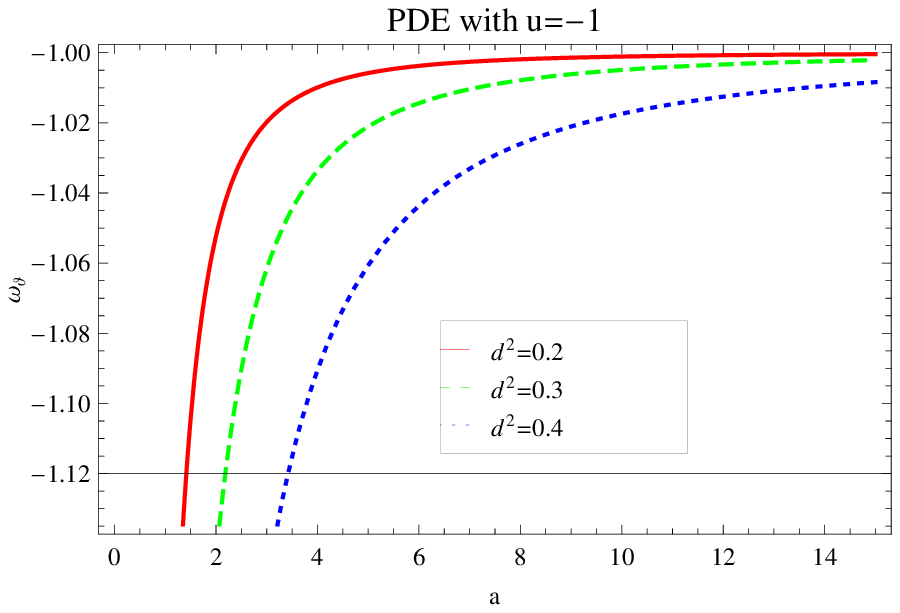,width=.50\linewidth}\caption{Plot of
$\omega_{\vartheta}$ versus $a$ for PDE in LQC with $u=-1$.}
\end{figure}
\begin{figure} \centering
\epsfig{file=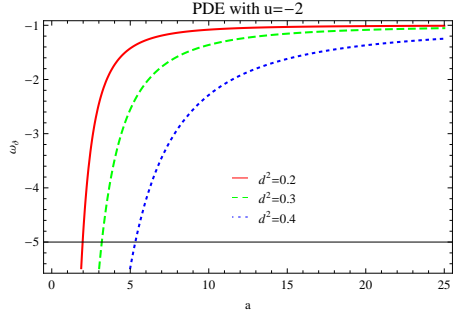,width=.50\linewidth}\caption{Plot of
$\omega_{\vartheta}$ versus $a$ for PDE in LQC with $u=-2$.}
\end{figure}

\begin{eqnarray*}
\omega_{\vartheta}&=&-1.13^{+0.24}_{-0.25}~~~~~~~\text{(Planck+WP+BAO)},\\
\omega_{\vartheta}&=&-1.09\pm0.17,~~~\text{(Planck+WP+Union 2.1)}\\
\omega_{\vartheta}&=&-1.13^{+0.13}_{-0.14},~~~~~~\text{(Planck+WP+SNLS)},\\
\omega_{\vartheta}&=&-1.24^{+0.18}_{-0.19},~~~~~~\text{(Planck+WP+$H_0$)},
\end{eqnarray*}
by implying different combination of observational schemes at $95\%$
confidence level. It can be seen from Figures \textbf{5-8} that the
EoS parameter also meets the above mentioned values for all cases of
interacting parameter which shows consistency of our results. The
above discussion shows that all the models provides fully support
the PDE phenomenon.

\subsection{The Square Speed of Sound}

In order to analyze the stability of PDE model in this scenario, we
extract the squared speed of sound which is given by
\begin{equation}\label{15}
\upsilon_{s}^2=\frac{\dot{p}}{\dot{\rho}}=\frac{p'}{\rho'},
\end{equation}
where pressure corresponds to PDE only. After some calculations, we
can obtain squared speed of sound as follows
\begin{eqnarray}\nonumber
\upsilon^2_{s}&=&\frac{1}{3}(-3-(a^{-3+3d^2}d^2\rho_{m0}H(a)^{-u})n^{-2}
+(a^{-3+3d^2}\rho_{m0}u(-un^2\\\nonumber &\times&H(a)^{-2 +
u}+2\rho_{c}(-2a^{-3+3d^2}d+\rho_{c}-6n^2H(a)^u)^{-1}))\\\nonumber
&\times&H^{-2}(a)-(a^{-3
+3d^2}d^2\rho_{m0}u^2-(3(-1+d^2)d^2H(a)^{4-u})n^{-2}\\\nonumber
&-&3(-1+d^2)u^2 n^2H(a)^{1 +
u}-a^{-3+3d^2}\rho_{m0}(-3+u)u^3n^4H(a)^{-3+2u}\\\nonumber
&+&(6a^6(-1+bs)u \rho_{c}
H(a)^3(\rho_{c}-6n^2H(a)^u))(2a^{3d^2}\rho_{m0}-a^3\alpha\\\nonumber
&+&6a^3 n^2 H(a)^u)^{-2}+(2a^{3d^2}d^2\rho_{m0}u\rho_{c}
H(a)^{2-u})(n^2(2a^{3d^2}\rho_{m0}-a^3\rho_{c}\\\nonumber
&+&6a^3n^2H(a)^u))^{-1}+(2a^{3 d^2}\rho_{m0}u^2\rho_{c}
n^2H(a)^{-1+u}(-(-4+u)(2a^{3 d^2}\rho_{m0}\\\nonumber &-&a^3
\rho_{c})+24a^3n^2H(a)^u))(2
a^{3d^2}\rho_{m0}-a^3\rho_{c}+6a^3n^2H(a)^u)^{-2}\\\nonumber
&+&(4a^{3+3d^2}\rho_{m0}u
\rho_{c}^2H(a)(2a^{3d^2}\rho_{m0}-a^3\rho_{c}+6a^3(1+u)n^2H(a)^u))(2a^{3d^2}\\\nonumber
&\times&\rho_{m0}-a^3\rho_{c}+6a^3n^2H(a)^u)^{-3})(au
H(a)^3(-un^2H(a)^{-2+u}
\\\nonumber
&+&(2\rho_{c})(-2a^{-3+3d^2}\rho_{m0}+\rho_{c}-6n^2H(a)^u)^{-1}))^{-1}).
\end{eqnarray}
The plots of squared speed of sound versus $a$ for three different
values of $d^2$ and four values of $u=2,~1,-1,-2$ is shown in
Figures \textbf{9-12}. It can be observed from Figures \textbf{9-10}
(for cases $u=2,~1$) that the squared speed of sound remains
negative for all cases of $d^2$ which exhibits the instability of
the PDE in LQC scenario. For the cases $u=-1,-2$ (Figures
\textbf{11-12}), it exhibits the stability of the present model for
all cases of $d^2$.

\begin{figure} \centering
\epsfig{file=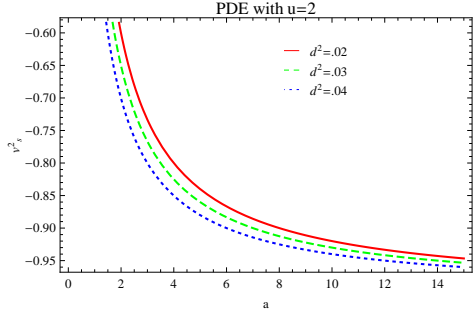,width=.50\linewidth}\caption{Plot of
$\upsilon^2_s$ versus $a$ for PDE in LQC with $u=2$.}
\end{figure}
\begin{figure} \centering
\epsfig{file=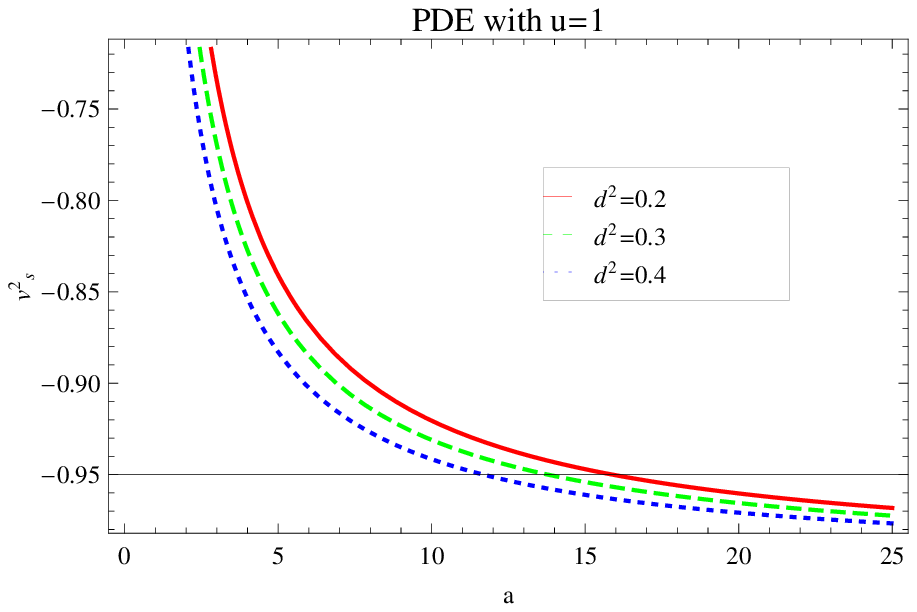,width=.50\linewidth}\caption{Plot of
$\upsilon^2_s$ versus $a$ for PDE in LQC with $u=1$.}
\end{figure}
\begin{figure} \centering
\epsfig{file=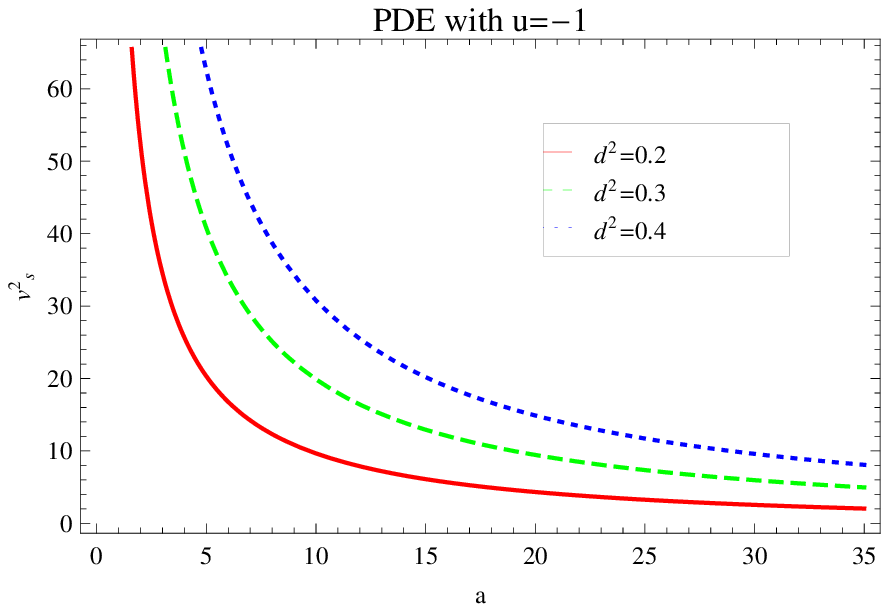,width=.50\linewidth}\caption{Plot of
$\upsilon^2_s$ versus $a$ for PDE in LQC with $u=-1$.}
\end{figure}
\begin{figure} \centering
\epsfig{file=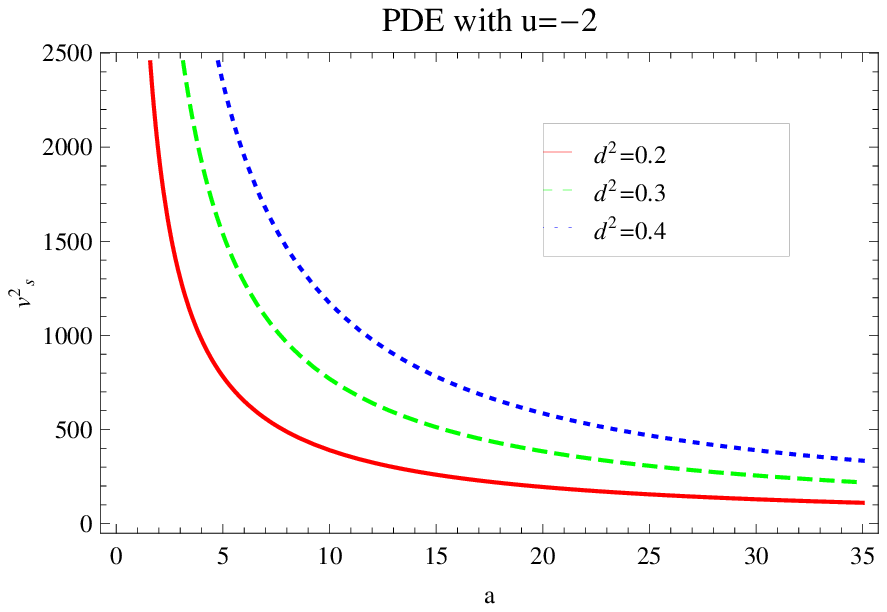,width=.50\linewidth}\caption{Plot of
$\upsilon^2_s$ versus $a$ for PDE in LQC with $u=-2$.}
\end{figure}

\section{Cosmological Planes in LQC}

Here, we will discuss the physical significance of cosmological
planes corresponding to PDE with Hubble horizon in LQC scenario.

\subsection{$\omega_{\vartheta}-\omega'_{\vartheta}$ Analysis}

Here, we find the regions on the
$\omega_{\vartheta}-\omega'_{\vartheta}$ plane
($\omega'_{\vartheta}$ represents the evolution of
$\omega_{\vartheta}$) as defined by Caldwell and Linder \cite{C43}
for models under consideration. The models can be categorized in two
different classes as thawing and freezing regions on the
$\omega_{\vartheta}-\omega'_{\vartheta}$ plane. The thawing models
describe the region $\omega'_{\vartheta}>0$ when
$\omega_{\vartheta}<0$ and freezing models represent the region
$\omega'_{\vartheta}<0$ when $\omega_{\vartheta}<0$. Initially, this
phenomenon was applied for analyzing the behavior of quintessence
model and found that the corresponding area occupied on the
$\omega_{\vartheta}-\omega'_{\vartheta}$ plane describes the thawing
and freezing regions. Differentiating $\omega_{\vartheta}$ with
respect to $x=\ln a$ and after some calculations, we obtain
\begin{eqnarray}\nonumber
\omega'_{\vartheta}&=&(3H(a)^4)^{-1}a^{-4+3d^2}\rho_{m0}(a^{-3+3d^2}d^4\rho_{m0}
u^2 -(3(-1+d^2)d^4H(a)^{4-u})\\\nonumber &\times&n^{-2}-3(-1+d^2)u^2
n^2H(a)^{1+u}-a^{-3+3d^2}\rho_{m0}(-3+u)u^3n^4H(a)^{-3+2u}
\\\nonumber &+&(6a^6(-1+d^2)u\rho_c
H(a)^3(\rho_c-6n^2H(a)^u))(2a^{3d^2}\rho_{m0}-a^3
\rho_c+6a^3\\\nonumber
&\times&n^2H(a)^u)^{-2}+(2a^{3d^2}d^4\rho_{m0}u\rho_c
H(a)^{2-u})(n^2(2a^{3d^2}\rho_{m0}-a^3\rho_c\\\nonumber
&+&6a^3n^2H(a)^u))^{-2}+(2a^{3d^2}\rho_{m0}u^2\rho_c
n^2H(a)^{-1+u}(-(-4+u)(2a^{3d^2}\rho_{m0}\\\nonumber &-&a^3 \rho_c)+
24a^3
n^2H(a)^u))(2a^{3d^2}\rho_{m0}-a^3\rho_c+6a^3n^2H(a)^u)^{-2}+(4a^{3+
3d^2}\\\nonumber &\times&\rho_{m0} u\rho_c^2H(a)(2a^{3
d^2}\rho_{m0}-a^3\rho_c+6a^3(1+u)n^2H(a)^u))(2a^{3
d^2}\rho_{m0}\\\nonumber &-&a^3\rho_c+6a^3n^2H(a)^u)^{-3}).
\end{eqnarray}
\begin{figure} \centering
\epsfig{file=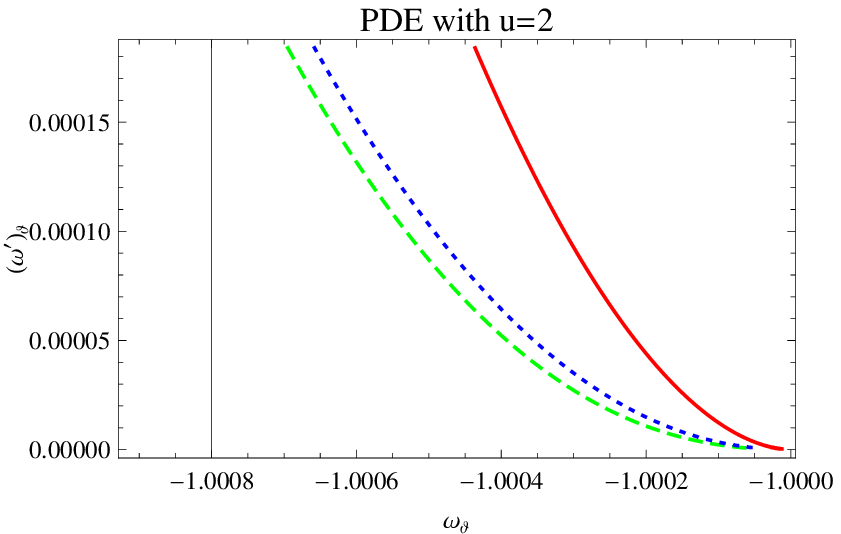,width=.50\linewidth}\caption{Plot of
$\omega_{\vartheta}-\omega'_{\vartheta}$ for PDE in LQC with $u=2$.}
\end{figure}
\begin{figure} \centering
\epsfig{file=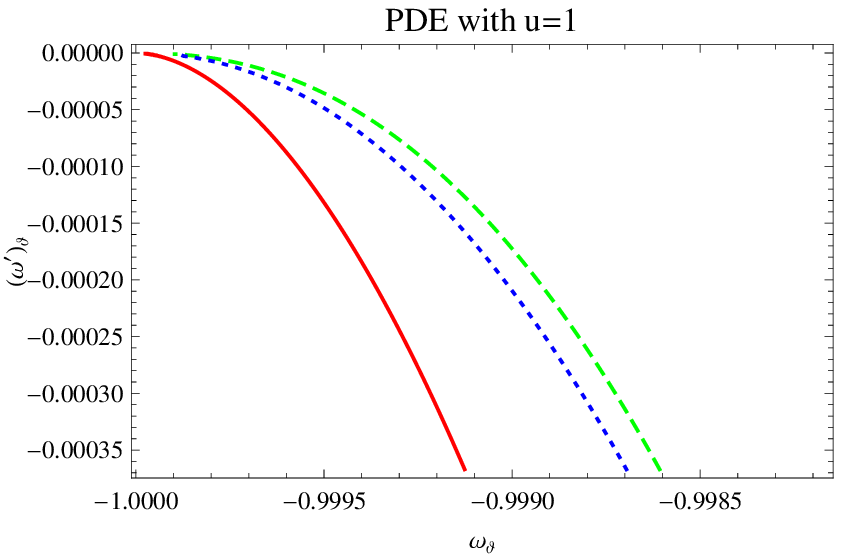,width=.50\linewidth}\caption{Plot of
$\omega_{\vartheta}-\omega'_{\vartheta}$ for PDE in LQC with $u=1$.}
\end{figure}
\begin{figure} \centering
\epsfig{file=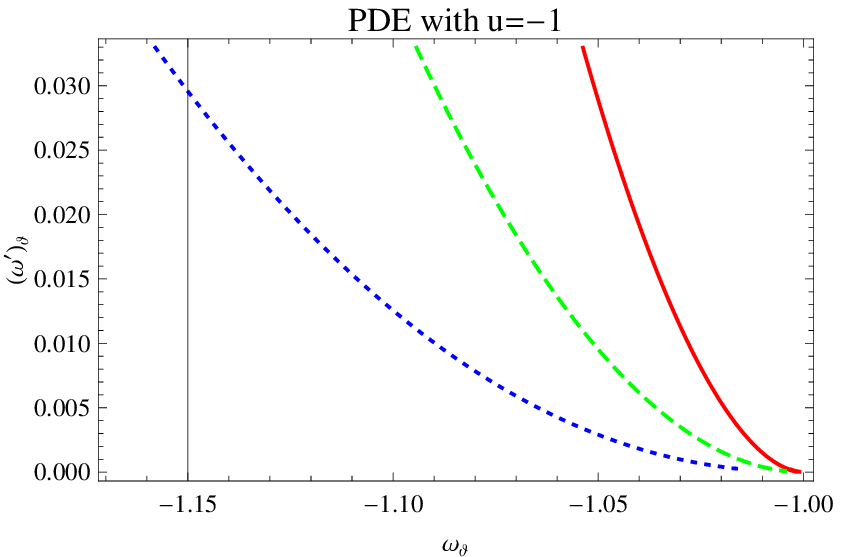,width=.50\linewidth}\caption{plot of
$\omega_{\Lambda}-\omega'_{\Lambda}$ for PDE in LQC with $u=-1$.}
\end{figure}
\begin{figure} \centering
\epsfig{file=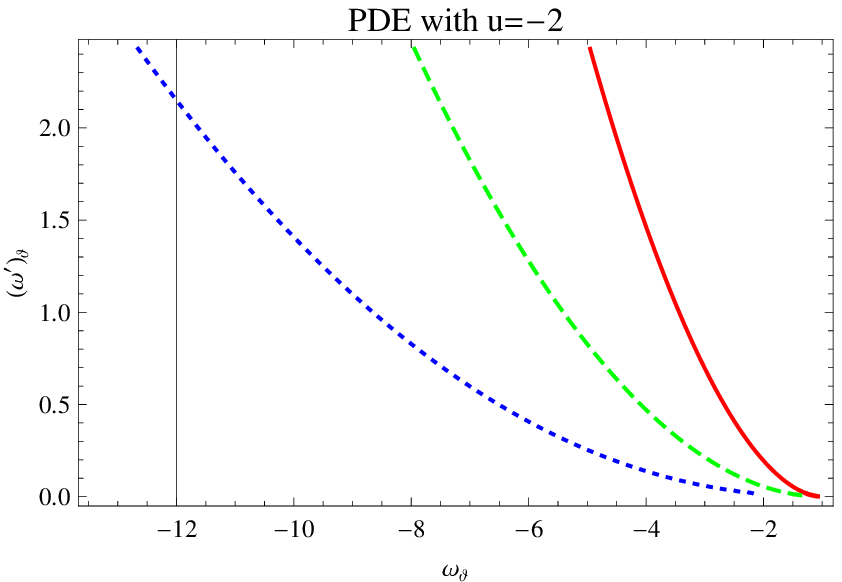,width=.50\linewidth}\caption{plot of
$\omega_{\Lambda}-\omega'_{\Lambda}$ for PDE in LQC with $u=-2$.}
\end{figure}

We also construct the $\omega_{\vartheta}-\omega'_{\vartheta}$ plane
for PDE model with different values of $u$ in LQC as shown in
Figures \textbf{13-16}. In all cases of $u$ and $d^2$, the
$\omega_{\vartheta}-\omega'_{\vartheta}$ plane corresponds to
$\Lambda$CDM limit, i.e.,
$(\omega_{\vartheta},\omega'_{\vartheta})=(-1,0)$. Also,
$\omega_{\vartheta}-\omega'_{\vartheta}$ plane shows thawing regions
for the cases $u=2,-1,-2$ (Figures \textbf{13,15,16}) and
corresponds to freezing region for the case $u=1$ (Figure
\textbf{14}). It has been developed the following constraints on
$w_{\vartheta}$ and $w'_{\vartheta}$ by Ade et al. \cite{Planck}:
\begin{eqnarray*}
\omega_{\vartheta}&=&-1.13^{+0.24}_{-0.25}~~~~~~~\text{(Planck+WP+BAO)},\\
\omega'_{\vartheta}&<&1.32,~~~~~~~~~~~~~~\text{(Planck+WP+BAO)}
\end{eqnarray*}
at $95\%$ confidence level. Also, other data with different
combinations of observational schemes such as (Planck+WP+Union 2.1)
and (Planck+WP+SNLS) favor the above constraints. In the present
case, the trajectories of $\omega'_{\vartheta}$ against
$\omega_{\vartheta}$ also meet the above mentioned values for all
cases of interacting parameter which shows consistency of our
results as shown in Figures \textbf{13-16}. Hence,
$\omega_{\vartheta}-\omega'_{\vartheta}$ plane provides consistent
behavior with the present day observations in all cases of $u$.

\subsection{Statefinder Parameters}

The statefinder parameters are defined as follows \cite{18}
\begin{eqnarray}\label{21}
r=\frac{\dddot{a}}{aH^3},\quad s=\frac{r-1}{3(q-\frac{1}{2})},
\end{eqnarray}
where $q$ is the deceleration parameter. These parameters are
dimensionless and possess the ability to explain the current
accelerated scenario. These parameters have geometrical diagnostic
due to their total dependence on the expansion factor. The
statefinders are useful in the sense that we can find the distance
of a given DE model from $\Lambda$CDM limit. The well-known regions
described by these cosmological parameters are as follows:
$(r,s)=(1,0)$ indicates $\Lambda$CDM limit, $(r,s)=(1,1)$ shows CDM
limit, while $s>0$ and $r<1$ represent the region of phantom and
quintessence DE eras. By following the papers \cite{N6,N7}, we can
obtained the following form of statefinders
\begin{eqnarray}\nonumber
r&=&1+\frac{1}{2}a^{-3+3d^2}\rho_{m0}(-u^2n^2H a^{-3+u}-(d^4
H(a)^{-u})n^{-2}+(2u\rho_{c})(H(a)\\\nonumber&\times&(-2a^{-3+3d^2}\rho_{m0}+\rho_{c}-6n^2
H(a)^u))^{-1})(-3-(a^{-3+3d^2}d^4\rho_{m0}
H(a)^{-u})\\\nonumber&\times&n^{-2}+(a^{-3+3d^2}\rho_{m0}u(-un^2H(a)^{-2
+u}+(2\rho_{c})(-2a^{-3+3d^2}\rho_{m0}+\rho_{c}\\\nonumber&-&6n^2H(a)^u)^{-1}))H(a)^{-1}-(3a^{3-3d^2}H(a)
\omega'_{\vartheta})(\rho_{m0}u(-un^2H(a)^{-2+u}\\\label{r2}&+&(2\rho_{c})(-2a^{-3+3d^2}\rho_{m0}+\rho_{c}-
6n^2H(a)^u)^{-1}))^{-1}).
\end{eqnarray}
and
\begin{eqnarray}\nonumber
s&=&\frac{1}{9}a^{-3+3d^2}\rho_{m0}(-u^2n^2H(a)^{-3 + u}-(d^4
H(a)^{-u})n^{-2}+(2u\rho_{c})(H(a)\\\nonumber&\times&(-2a^{-3 +
3d^2}\rho_{m0}+\rho_{c}-6n^2
H(a)^u))^{-1})(-3-(a^{-3+3d^2}d^4\rho_{m0}H(a)^{-u})\\\nonumber&\times&n^{-2}+(a^{-3
+ 3 d^2}\rho_{m0} u(-un^2H(a)^{-2 +
u}+(2\rho_{c})(-2a^(-3+3d^2)\rho_{m0}\\\nonumber&+&\rho_{c}-
6n^2H(a)^u)^{-1}))H(a)^{-1}-(3a^{3-3d^2}H(a)\omega'_{\vartheta})(\rho_{m0}
u(-un^2H(a)^{-2+u}\\\label{r3}&+&(2\rho_{c})(-2a^{-3+3d^2}\rho_{m0}+\rho_{c}-
6n^2H(a)^u)^{-1}))^{-1}).
\end{eqnarray}
\begin{figure} \centering
\epsfig{file=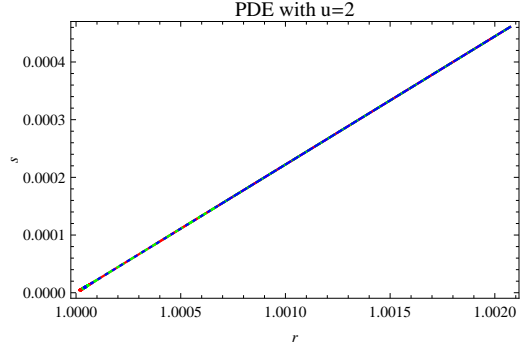,width=.50\linewidth}\caption{Plot of $r-s$ for
for PDE in LQC with $u=2$.}
\end{figure}
\begin{figure} \centering
\epsfig{file=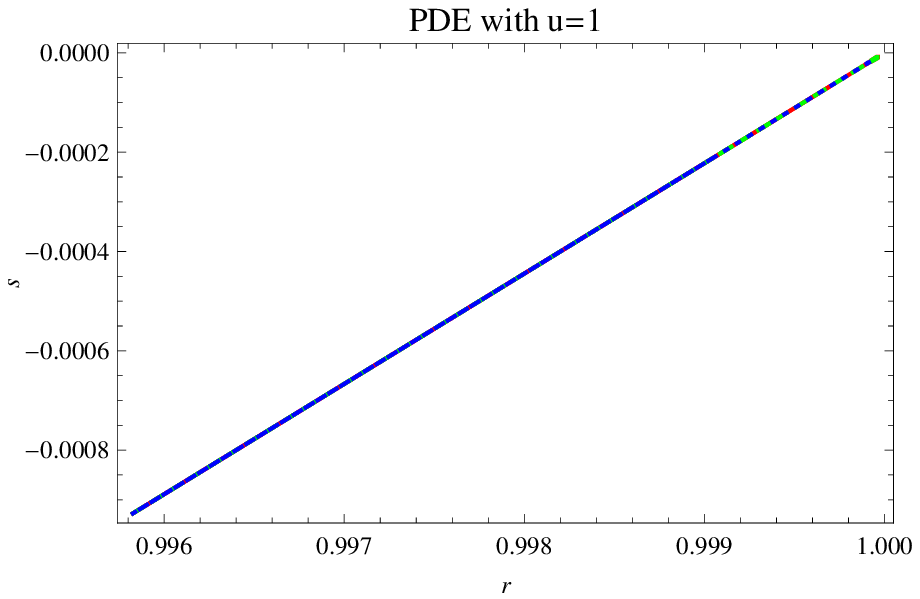,width=.50\linewidth}\caption{$r-s$ for PDE for
PDE in LQC with $u=1$.}
\end{figure}
\begin{figure} \centering
\epsfig{file=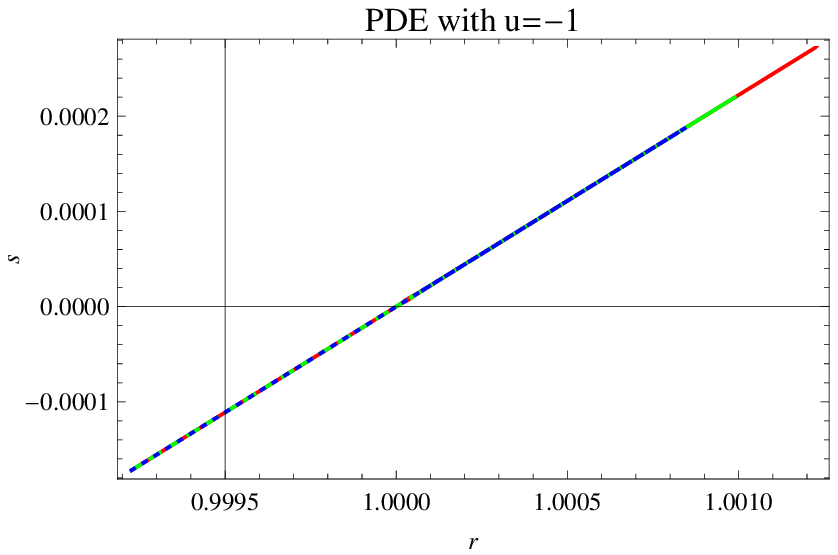,width=.50\linewidth}\caption{$r-s$ for PDE in
LQC with $u=-1$.}
\end{figure}
\begin{figure} \centering
\epsfig{file=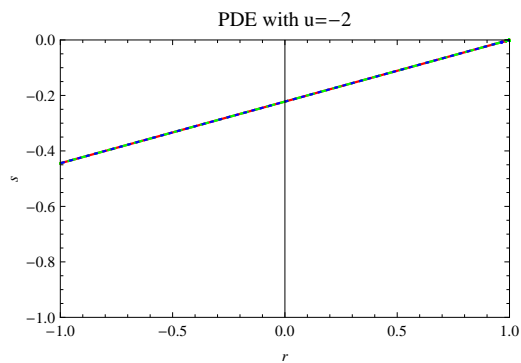,width=.50\linewidth}\caption{$r-s$ for PDE in
LQC with $u=-2$.}
\end{figure}

We also develop $r-s$ planes corresponding to the present
cosmological scenario for different values of $u$ as shown in
Figures \textbf{17-20}. The $r-s$ corresponds to $\Lambda$CDM limit
for all cases of $u$.

\section{Concluding Remarks}

We have considered the framework of interacting PDE with Hubble
horizon in LQC framework. The main motivation of this work is to
analyze the cosmological scenario as well as construction of
possible constraints of PDE parameter $u$ where it fulfill the PDE
phenomenon. For this purpose, we have constructed Hubble parameter,
EoS parameter, squared speed of sound,
$\omega_{\vartheta}-\omega'_{\vartheta}$ and $r-s$ planes,
numerically. We have discussed these parameters corresponding to
four values of $u=2,~1,-1,-2$ and three value of
$d^2=0.2,~0.3,~0.4$. We have observed that the trajectories of
Hubble parameter $H(a)$ for all cases of $u$ attains the values
approximately to $74^{+0.005}_{-0.005}$ (Figures \textbf{1-4}).
These obtained range of $H(a)$ shows consistency with the
observational values such as $H_0=73.8\pm2.4$ \cite{Riess} and
$H_0=74.3 \pm 1.5$ \cite{Freedman}.

Moreover, the EoS parameter also shows consistency with the present
day observations. For instance, the trajectory of
$\omega_{\vartheta}$ exhibits the ranges $-1-0.0050$ and
$-1+0.00005$ for the cases $u=2,~1$ as shown in Figures
\textbf{5-6}. For $u=-1,-2$ (Figures \textbf{7-8}), the EoS
parameter lies in the ranges $(-1.12,-1)$ and $(-5,-1)$,
respectively. These constraints on EoS parameter compatible with the
constraints as obtained by Ade et al. \cite{Planck} (Planck data)
which is given as follows:
\begin{eqnarray*}
\omega_{\vartheta}&=&-1.13^{+0.24}_{-0.25}~~~~~~~\text{(Planck+WP+BAO)},\\
\omega_{\vartheta}&=&-1.09\pm0.17,~~~\text{(Planck+WP+Union 2.1)}\\
\omega_{\vartheta}&=&-1.13^{+0.13}_{-0.14},~~~~~~\text{(Planck+WP+SNLS)},\\
\omega_{\vartheta}&=&-1.24^{+0.18}_{-0.19},~~~~~~\text{(Planck+WP+$H_0$)}.
\end{eqnarray*}
The above constraints has been obtained by implying different
combination of observational schemes at $95\%$ confidence level.

It can also be observed from Figures \textbf{9-10} (for cases
$u=2,~1$) that the squared speed of sound remains negative for all
cases of $d^2$ which exhibits the instability of the PDE in LQC
scenario. For the cases $u=-1,-2$ (Figures \textbf{11-12}), it
exhibits the stability of the present model for all cases of $d^2$.
We have also observed that the
$\omega_{\vartheta}-\omega'_{\vartheta}$ plane corresponds to
$\Lambda$CDM limit, i.e.,
$(\omega_{\vartheta},\omega'_{\vartheta})=(-1,0)$ in all cases of
$u$ and $d^2$. Also, $\omega_{\vartheta}-\omega'_{\vartheta}$ plane
shows thawing regions for the cases $u=2,-1,-2$ (Figures
\textbf{13,15,16}) and corresponds to freezing region for the case
$u=1$ (Figure \textbf{14}). It has been developed the following
constraints on $w_{\vartheta}$ and $w'_{\vartheta}$ by Ade et al.
\cite{Planck}:
\begin{eqnarray*}
\omega_{\vartheta}&=&-1.13^{+0.24}_{-0.25}~~~~~~~\text{(Planck+WP+BAO)},\\
\omega'_{\vartheta}&<&1.32,~~~~~~~~~~~~~~\text{(Planck+WP+BAO)}
\end{eqnarray*}
at $95\%$ confidence level. Also, other data with different
combinations of observational schemes such as (Planck+WP+Union 2.1)
and (Planck+WP+SNLS) favor the above constraints. In the present
case, the trajectories of $\omega'_{\vartheta}$ against
$\omega_{\vartheta}$ also meet the above mentioned values for all
cases of interacting parameter which shows consistency of our
results as shown in Figures \textbf{13-16}. Hence,
$\omega_{\vartheta}-\omega'_{\vartheta}$ plane provides consistent
behavior with the present day observations in all cases of $u$.
Also, the $r-s$ corresponds to $\Lambda$CDM limit for all cases of
$u$. Finally, it is remarked that all the cosmological parameters in
the scenario of LQC with PDE shows compatibility with the current
observations.


\begin{thebibliography}{43}

\bibitem{1aa} Perlmutter, S. et al.: Astrophys. J. \textbf{517}(1999)565;
Caldwell, R.R. and Doran, M.: Phys. Rev. D \textbf{69}(2004)103517;
Koivisto, T. and Mota, D.F.: Phys. Rev. D \textbf{73}(2006)083502;
Daniel, S.F.: Phys. Rev. D \textbf{77}(2008)103513; Fedeli, C.,
Moscardini, L. and Bartelmann, M.: Astron. Astrophys.
\textbf{500}(2009)667.

\bibitem{1a} Peebles, P.J.E.: Rev. Mod. Phys. \textbf{75}(2003)559.

\bibitem{2} Kamenshchik, A.Y., Moschella, U. and Pasquier, V.: Phys. Lett. B \textbf{511}(2001)265;
Bento, M.C., Bertolami, O. and Sen, A.A.: Phys. Rev. D
\textbf{66}(2002)043507; Zhang, X., Wu, F.Q. and Zhang, J.: JCAP
\textbf{01}(2006)003.

\bibitem{3} Hsu, S.D.H.: Phys. Lett. B \textbf{594}(2004)13.

\bibitem{4} Li, M.: Phys. Lett. B \textbf{603}(2004)1.

\bibitem{5} Cai, R.G.: Phys. Lett. B \textbf{660}(2008)113.

\bibitem{6} Karami, K., Ghaffari, S. and Fehri, J.: Eur. Phys. J. C
\textbf{64}(2009)85.

\bibitem{N5} Wei, H.: Class. Quantum Grav. \textbf{29}, 175008
(2012).

\bibitem{N6} Sharif, M. and Jawad, A.: Eur. Phys. J. C \textbf{73}, 2382
(2013).

\bibitem{N7} Sharif, M. and Jawad, A.: Eur. Phys. J. C \textbf{73}, 2600
(2013).

\bibitem{suss} Susskind, L.: J. Math. Phys. \textbf{36}(1995)6377.

\bibitem{cohen} Cohen, A., Kaplan, D. and Nelson, A.: Phys. Rev. Lett. \textbf{82}(1999)4971.

\bibitem{5a} Li, M.: Phys. Lett. B \textbf{603}(2004)1.

\bibitem{5b} Wei, H. and Cai, R.G.: Phys. Lett. B \textbf{660}(2008)113.

\bibitem{5c} Gao, C., Chen, X. and Shen, Y.G.: Phys. Rev. D
\textbf{79}(2009)043511; Granda, L. and Oliveros, A.: Phys. Lett. B
\textbf{669}(2008)275; Chen, S. and Jing, J.: Phys. Lett. B
\textbf{679}(2009)144.

\bibitem{shar} Sharif, M. and Jawad, A.: Eur. Phys. J. Plus \textbf{129}, 15
(2014); Lobo, F.S.N.: Phys. Rev. D \textbf{71}(2005)124022; Lobo,
F.S.N.: Phys. Rev. D \textbf{71}(2005)084011;; Sushkov, S.: Phys.
Rev. D \textbf{71}(2005)043520.

\bibitem{30} Sharif, M. and Jawad, A.: Int. J. Mod. Phys. D \textbf{22}, 1350014 (2013);
Martin-Moruno, P.: Phys. Lett. B \textbf{659}(2008)40; Jamil, M.,
Rashid, M.A. and Qadir, A.: Eur. Phys. J. C \textbf{58}(2008)325;
Babichev, E. et al.: Phys. Rev. D \textbf{78}(2008)104027; Jamil,
M.: Eur. Phys. J. C \textbf{62}(2009)325; Jamil, M. and Qadir, A.:
Gen. Rel. Grav. \textbf{43}(2011)1069; Bhadra, J. and Debnath, U.:
Eur. Phys. J. C \textbf{72}(2012)1912.

\bibitem{9} Gonzalez, J.A. and Guzman, F.S.: Phys. Rev. D \textbf{79}(2009)121501.

\bibitem{10} Sun, C.Y.: Commun. Theor. Phys. \textbf{52}(2009)441.

\bibitem{11} Harada, T., Maeda, H. and Carr, B.J.: Phys. Rev. D \textbf{74}(2006)024024;
Akhoury, R., Gauthier, C.S. and Vikman, A.: JHEP 03(2009)082.

\bibitem{S1} Cai, Y-F., et al.: Phys. Reports \textbf{493}(2010)1.

\bibitem{S2} Saridakis, E.N.: Nucl. Phys. B {\bf 819}(2009)116.

\bibitem{S3} Gupta, G., Saridakis, E.N. and Sen, A.A.: Phys. Rev. D \textbf{79}(2009)123013.

\bibitem{S4} Setare, M.R. and Saridakis, E.N.: JCAP
\textbf{0903}(2009)002.

\bibitem{S4} Setare, M.R. and Saridakis, E.N.: Phys. Lett. \textbf{B}
671(2009)331.

\bibitem{S5} Saridakis, E.N., Gonzalez-Diaz, P.F. and Siguenza, C.L.: Class. Quant. Grav.
\textbf{26}(2009)165003.

\bibitem{S6} Saridakis, E.N.: Phys. Lett. B \textbf{676}(2009)7.

\bibitem{S7} Saridakis, E.N.: Phys. Lett. B \textbf{660}(2008)138.

\bibitem{S8} Saridakis, E.N.: Phys.Lett.B661:335-341,2008.

\bibitem{S9} Setare, M.R. and Saridakis, E.N.: Phys. Lett. B
\textbf{671}(2009)331.

\bibitem{N47} Sharif, M. and Rani, S.: J. Exp. Theor. Phys. (to appear,
2014).

\bibitem{N48} Chattopadhyay, S., Jawad, A., Momeni, D. and Myrzakulov,
R.:Astrophys. Space Sci. \textbf{353}, 279 (2014).

\bibitem{N49} Jawad, A.: Astrophys. Space Sci. (to appear, 2014).

\bibitem{L22} Bojowald, M.: Living Rev. Rel. \textbf{8}(2005)11.

\bibitem{L23} Ashtekar, A., Bojowald, M. and
Lewandowski, J.: Adv. Theor. Math. Phys. \textbf{7}(2003)233.

\bibitem{L24} Ashtekar, A.: AIP Conf. Proc. \textbf{861}(2006)3.

\bibitem{L25} Rovelli, C.: Living Rev. Rel. \textbf{1}(1998)1.

\bibitem{L26} Ashtekar, A. and Lewandowski, J.: Class. Quant. Grav. \textbf{21}, R53
(2004).

\bibitem{L27} Rovelli, C.: Quantum Gravity, Cambridge University Press,
Cambridge (2004).

\bibitem{L27a} Wu, P. and Zhang, S. N., 2008, JCAP 06, 007.

\bibitem{L27b} Chen, S., Wang, B. and Jing, J., 2008, Phys. Rev. D 78, 123503.

\bibitem{L29} Jamil, M. and Debnath, U., 2011, Astrophys Space Sci. 333, 3. [27]

\bibitem{L30} Fu, X., Yu, H. and Wu, P., 2008, Phys. Rev. D 78, 063001.

\bibitem{L31} Chakraborty, S., Debnath, U. and Ranjit, C.: Eur. Phys. J. C
\textbf{72}(2012)2101.

\bibitem{20a} Pavon, D. and Zimdahl, W.: Phys. Lett. B \textbf{628}(2005)206;
Zimdahl, W. and Pavon, D.: Class. Quantum Grav.
\textbf{24}(2007)5461.

\bibitem{20b} Dur$\acute{a}$n, I, Pav$\acute{o}$n, D. and Zimdahlb,
W.: JCAP \textbf{07}(2010)018.

\bibitem{20c} Gong, Y. and Li, T.: Phys. Lett. B \textbf{683}(2010)241.

\bibitem{12a} Sheykhi, A.: Phys. Rev. D \textbf{84}(2011)107302.

\bibitem{Planck} Ade, P.A.R., et al.: arXiv:1303.5076.

\bibitem{Riess} Riess, A. G., et al.: Astrophys. J.\textbf{730}(2011)119.

\bibitem{Freedman} Freedman, W. L., et al.: Astrophys. J.\textbf{758}(2012)24.

\bibitem{C43} Caldwell, R.R. and Linder, E.V.: Phys. Rev. Lett. \textbf{95}(2005)141301.

\bibitem{18} Sahni, V. et al.: JETP Lett. \textbf{77}(2003)201.

\end{thebibliography}
\end{document}